\documentclass[prb,twocolumn,showpacs,8pt]{revtex4}
\usepackage{dcolumn}
\usepackage{bm}
\usepackage[all]{xy}
\def\bea{\begin{eqnarray}}
\def\eea{\end{eqnarray}}
\usepackage{graphicx,amsbsy,color,amssymb,amsmath}
\usepackage[all]{xy}

\newcommand{\ket}[1]{\ensuremath{|{#1}\rangle}}

\renewcommand{\Re}{\ensuremath{\text{Re}}}
\renewcommand{\Im}{\ensuremath{\text{Im}}}

\usepackage{amsmath}
\usepackage{color}
\usepackage{soul}

\begin{document}
 \title{Theory of  Superconductivity in the Cuprates}
\author{Vivek Aji, Arkady Shekhter and C. M. Varma}
\affiliation{Department of Physics and Astronomy, University of
California, Riverside, California 92521}
\begin{abstract}
The quantum critical fluctuations of the time-reversal breaking order parameter which is observed in the pseudogap regime of the Cuprates are shown to couple to the lattice equivalent of the local angular momentum of the fermions. Such a coupling favors scattering of fermions through angles close to $\pm \pi/2$  which is unambiguously shown to promote d-wave pairing. The right order of magnitude of both $T_c$ and the normalized zero temperature gap $\Delta/T_c$ are calculated using the same fluctuations which give the temperature, frequency and momentum dependence of the the anomalous normal state properties for dopings near the quantum-critical value and with two parameters extracted from fit to such experiments. \end{abstract}
\maketitle

\section{Introduction}
The objective of a microscopic theory of the phenomena in 
Cuprates ought to be to derive their {\it universal} properties, in all the parts of their phase diagram, based on a single set of ideas and with a consistent set of calculations on a well-defined model.  In particular, since superconductivity is an instability of the normal state of a metal, it is necessary that the same theory which seeks to explain high temperature superconductivity in the Cuprates also explain their remarkable normal states. To be convincing, the theory should also lead to unique predictions which can be tested in experiments. 

Towards these goals, one particular approach to the theory has so far achieved the following  :\\
 (1) Starting from the three-orbital model with on-site and nearest neighbor repulsions, it was predicted that the pseudogap state occurs through a phase transition to a new state of matter with spontaneous orbital currents without changing translational symmetry. The transition temperature $T^*(x) \to 0$ for $x \to x_c$, defining  a quantum critical point at $x=x_c$. The loop current order parameter \cite{cmv} has by now been observed in the pseudogap region of four distinct families of Cuprates \cite{kaminski, fauque, mook, greven, pc}. \\
(2) The quantum critical fluctuations (QCF) of the observed order have been derived \cite{aji-cmv} and shown to have the spectrum with $\omega/T$ scaling and spatial locality, which was introduced phenomenologically long ago \cite{mfl} to explain the observed normal state anomalies \cite{physreport} and predict the single-particle spectra in the 'Marginal Fermi-liquid' region around $x=x_c$. Using ARPES data the parameters of the QCF's and their coupling strength to fermions have also been determined \cite{zhu/waterfall}.

In this paper, we show that the derived QCF couple to (the lattice equivalent of) the local orbital angular momentum of the fermions. This is a natural generalization of the coupling of the spin-angular momentum of the fermions to collective spin-fluctuations \cite{msrcmv}. We derive the momentum-dependence of the coupling of the QCF to fermions of the conduction band and show from microscopic theory as well as symmetry considerations that their exchange leads to an attractive d-wave pairing. Using the parameters extracted from the quantitative fit to the normal state anomalies and Angle-Resolved Photoemission Spectra (ARPES), we obtain the right order of magnitude of $T_c$ and $\Delta$ for superconductivity in the region dominated by the QCF. The principal findings of this paper can be tested in detailed analysis of ARPES data in the superconducting state using the generalization of the McMillan-Rowell procedure for s-wave superconductors.
\subsection{Plan of this Paper}
In order to present the new results of this paper, it is useful to briefly recapitulate earlier work upon which it builds. In Sec. II, we summarize (i) the microscopic model, (ii) the derivation of the Loop-current order based on it, and (iii) the quantum statistical mechanical model for the quantum fluctuations of the order parameter. As shown earlier, lattice anisotropy is irrelevant in the fluctuation regime and the spectra of the fluctuations is obtained from the solution of the dissipative quantum xy or rotor model in the continuum. 
In Sec. III, we present the coupling of the fermions to the fluctuations based on general symmetry considerations and show that in the continuum, the fluctuations of the angular momentum of the rotors couple to the local angular momentum operator of the fermions. In Sec. IV, we derive, through microscopic calculations, the coupling of the fermions to the fluctuations in the lattice model and show that it is a generalization of the continuum model to take into account the lattice anisotropy. Some technical details of the lattice calculations are given in the Appendices. In Sec. V, we present the vertex for superconductivity derived from the couplings in Secs. III and IV and prove that only d-wave pairing is possible in the model. In Sec. VI, we derive parameters of the model from the fit to the normal state spectral function $A({\bf k}, \omega)$ of the fermions and estimate $T_c$ and the superconducting gap $\Delta$ based on generalization of the Eliashberg theory to d-wave superconductors. We also discuss the limit of validity of the Eliashberg theory in the present context.

\section{Summary of work leading up to the present work}
\subsection{Microscopic Model} 

\begin{figure}[ht!]
  \centerline{
  \includegraphics[width=0.5\columnwidth]{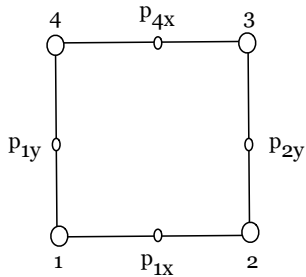}}
  \caption{ The unit-cell chosen for the two-dimensional model for Cuprates used for calculations in this paper. The labeling used to denote the Cu orbitals and the oxygen orbitals will be used throughout the paper.}
   \label{fig:unitcell}
\end{figure}

The relevant microscopic model for the loop-current order (LCO) in the Cuprates is the three orbital model \cite{vsa} with local {\it and} finite range interactions. A unit-cell with a Cu and two Oxygen orbitals per unit-cell and with labelling used in this paper is shown in fig.\ref{fig:unitcell}.
The nearest neighbor copper-oxygen interactions,
\begin{eqnarray}
\label{nn-int} H_{nn} = \sum_{\langle{R,R^{\prime}}\rangle} &&V_{pd} n_{d}(R)[n_{p_x} 
(R^{\prime})+n_{p_y}(R^{\prime}) \\ \nonumber
&&+ V_{pp} n_{p_x}(R)n_{p_y}(R^{\prime})].
\end{eqnarray}
play a crucial role in deriving the LCO. Here $ n_{d}(R), n_{p_{x,y}}(R')$ are the charge operator for the d-orbital on Cu site and the $p_{x,y}$-orbital on the oxygen sites at the sites $R$ and $R'$ respectively. Only neighboring Cu and Oxygen sites on the 8 links $(R,R')$ per each unit cell are summed. In the metallic state, the local interactions ($U's$) are assumed to only renormalize the kinetic energy parameters, unlike in the insulating-AFM state close to half-filling. 
The possible
novel changes in symmetry \cite{cmv} are seen by re-expressing Eq.~(\ref{nn-int}) using the operator identity,
\begin{align} \label{opident}
&2a_{\sigma}^{\dagger}(R)a_{\sigma}(R) \; b_{\sigma'}^{\dagger}(R^{\prime}) b_{\sigma'}(R^{\prime}) \notag\\
&\qquad=- \; |{\mathcal{O}}_{\sigma\sigma'}(R,R^{\prime})|^2 + a_{\sigma}^{\dagger}(R) a_{\sigma}(R) + b_{\sigma'}^{\dagger}(R^{\prime})b_{\sigma'}(R^{\prime}), \notag \\
&{ \mathcal{O}}_{\sigma\sigma'}(R,R^{\prime}) \equiv i a_{\sigma}^\dagger(R) b_{\sigma'}(R') + h.c.
\end{align}
${\bf \mathcal{O}}(R,R^{\prime}) \equiv \sum_{\sigma} {\mathcal{O}}_{\sigma\sigma}(R,R^{\prime})$ is proportional to the current operator on the link between site $i$ and $i^{\prime}$. 

Suppose an expectation value $<{\bf \mathcal{O}}(R,R^{\prime})> \equiv x(R,R')$ were shown to exist. Given a kinetic energy coefficient $t(R,R')$ in the bare Hamiltonian, this is equivalent to an effective kinetic energy operator on the link with a complex coefficient $(=t(R,R') + ix(R,R'))$. This amounts to a vector potential $\phi(R,R')=\arctan(x/t)$ on the link $(R,R')$. Gauge invariant combinations of vector potentials always form closed loops and correspond to flux in the area formed by the closed loops.
Such Gauge invariant combinations of $\phi_i(R,R')$ within a unit-cell $i$ are organized in to irreducible representations $\mu$ of the point group symmetry of the lattice:
\bea
\label{irred-rep-flux}
\tilde{L}_{i,\mu} \equiv \sum_{(R,R')} \phi_{i,\mu}(R,R').
\eea
As may be seen from Fig. (\ref{fig:unitcell}) and Fig. (\ref{fig:5fluxes}), there are 5 closed loops which can be formed in a unit-cell through connecting nearest neighbor Cu-O and O-O links. This is consistent with the vector potentials on each of the 8 links per unit-cell and the three lattice points per unit-cell at which independent gauge transformations can be made.  The algebraic representation as well as their flux patterns of the 5 varieties of the $\mu$ are exhibited in Appendix \ref{fluxes}. In the ground state in the pseudogap phase, one of these symmetries (with two-dimensional representation ${\bf E_g}$) is realized {\it globally} in the experiments and also found to be the lowest energy state in mean-field or better calculations \cite{weber}. There are differences in the experimental results from the prediction of the two-dimensional model. However, the spatial symmetry and the number of allowed configurations per unit-cell in the more general three dimensional model \cite{weber} are not changed from those given by the two-dimensional model \cite {shehter-cmv}. 

\begin{figure}[ht!]
  \centerline{
  \includegraphics[width=0.7\columnwidth]{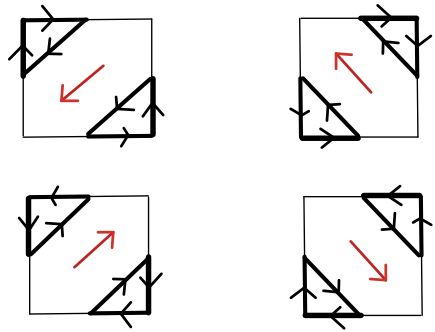}}
  \caption{ The four domains of the circulating current phase are shown. They may be specified by the four orientations of a vector ${\bf L}$ shown in red.}
   \label{fig:domain}
\end{figure}

The observed long-range ordered state is described by a time-reversal odd polar-vector ${\bf L}$ which has four possible orientations. 
These four domains are shown in fig.\ref{fig:domain}.  As discussed earlier \cite{shehter-cmv}, the generalization to the three dimensional solid including apical oxygens does not change the symmetry classification or the number of possible domains of the state. The most important fluctuations in the model in the quantum-critical regime are due to the transitions between the four configurations possible in this regime in each unit-cell. To describe the fluctuations,  we must in the minimal description retain the term which causes transitions among the four local configurations.  

\subsection{Effective Hamiltonian}

The complete Hamiltonian obtained using the identities in Eq. (\ref{opident}, {\ref{irred-rep-flux}}) is explicitly derived in appendix (\ref{loopham}). Limiting ourselves to the most important fluctuations, the derived effective Hamiltonian from which earlier results and the results in this paper are derived is
\begin{equation}\label{eqn:mfham}
H = K.E. - {(V_{pd})\over {16}} \left(\sum_{i,\ell}\tilde{\textbf{ L}}_{i,\ell}\cdot \tilde{\textbf{ L}}_{i,\ell} + |\tilde{\textbf{L}}_{z,i}|^2 \right),
\end{equation}
where
 \begin{eqnarray} 
 \tilde{\textbf{L}}_{i} = \left(\tilde{\textbf{L}}_{i,x'},\tilde{\textbf{L}}_{i,y'}\right)
\end{eqnarray}
The labels $i,i'$ will denote the unit-cells.
K.E. is the kinetic energy operator which is explicitly written down in appendix \ref{KE} in terms of two parameters $\tilde{t}_{pd}, \tilde{t}_{pp}$, which are respectively the effective hopping parameters (renormalized due to on-site repulsions) between the nearest neighbor Cu-O and O-O sites respectively. $\ell$ stands for $\hat{\bf x'},\widehat{\textbf{y'} }= ({\bf x \pm y})/\sqrt{2}$. $\tilde{\textbf{ L}}_{i,\ell}$ and $\tilde{\textbf{L}}_{z,i}$ are given in terms of the fermion operators in the Appendix \ref{fluxes}. The operators $\tilde{{\bf L}}_{z,i}$ introduce quantum fluctuations in the model through transitions between the four possible local configurations of the {\it local} order parameter ${\bf L}$, as derived in appendix (\ref{rotation}) and discussed below.
  
\noindent The operators  $\tilde{\textbf{L}}_i, \tilde{\textbf{L}}_{z,i}$ are expressed as the sum of a collective part 
${\textbf{L}}_i, {\textbf{L}}_{z,i}$ and a residual incoherent fermionic part ${\textbf{L}}_{fi},  {\textbf{L}}_{z,fi}$ using the Hubbard-Stratanovich or equivalent methods:
\begin{eqnarray}
\label{coll-ferm-sep}
\tilde{\textbf{L}}_i &=&  {\textbf{L}}_i + {\textbf{L}}_{fi} \\ \nonumber
\tilde{\textbf{L}}_{z,i} &=& {\textbf{L}}_{z,i} + {\textbf{L}}_{z,fi}.
\end{eqnarray}

 
The mean-field Hamiltonian consists of the kinetic energy Eq.(\ref{ke}) and the second term of Eq.(\ref{eqn:mfham}).
\noindent Integrating over the fermions generated a mean-field free-energy $F_{mf}(\left<{\bf L}\right>)$ \cite{cmv}. Minimizing this gave the stable long range order with an order parameter
 $< {\textbf{L}}_{i}> \equiv <{\bf \tilde{L}}_i>$ for all $i$.   
 
In this paper we are concerned with the coupling of the quantum critical fluctuations of the order parameter. The effective Hamiltonian for such fluctuations is generated from eqn. (\ref{eqn:mfham}) using the substitutions  eqn.\ref{coll-ferm-sep}. The fermion operators in the bilinear terms ${\textbf{L}}_i {\textbf{L}}_{fi} + {\textbf{L}}_j {\textbf{L}}_{fj} $  are eliminated by integrating over the propagator $\left<{\bf L}^+_{fi} {\bf L}_{fj}\right>$ in the standard manner to generate a coupling between the collective variables ${\textbf{L}}_i$ and ${\textbf{L}}_j$. Keeping coupling only between nearest neighbor cells this generates the fluctuation Hamiltonian,
\bea
H_{fl}= \sum_{(ij)} {\textbf{L}}_i ~\mathcal{J} ~{\textbf{L}}_j +  \frac{V_{pd}}{16}|{\textbf{L}}_{z,i}|^2.
\eea
Here $\mathcal{J}$ is in general a second rank tensor whose components depend on the orientations of ${\textbf{L}}_i$ and ${\textbf{L}}_j$. Only its order of magnitude can be estimated and is of $O(10^{-2} V^2/t )$, where $V$ is $V_{pd} \approx V_{pp}$ and $t$ are the kinetic energy parameters \cite{sudbo, shehterpc}. 

 \section{Quantum Fluctuations} 
The quantum model is specified in terms of operators ${\bf L}_i=e^{i\hat{\theta}_i}$, whose
eigenstates are the four angles $\theta_i$ in each cell $i$ depicted in Fig. (\ref{fig:domain}) :
\begin{equation}
{\bf L}_i\left|\theta_i\right\rangle=e^{i \theta_i}\left|\theta_i\right\rangle\,
\end{equation}
 \noindent Given the symmetries of the four domains, the angles correspond to
\begin{eqnarray}
\label{theta}
\left|\theta = \pi/4\right> &=& \left|\hat{x}+\hat{y}\right> \nonumber  \\
\left|\theta = 3\pi/4\right> &=& \left|-\hat{x}+\hat{y}\right> \nonumber  \\
\left|\theta = 5\pi/4\right> &=& \left|-\hat{x}-\hat{y}\right> \nonumber  \\
\left|\theta = 7\pi/4\right> &=& \left|\hat{x}-\hat{y}\right>
\end{eqnarray}

The low-energy quantum fluctuations of the order parameter form current loops of all sizes and shapes and varying in time. They are generated by the elementary process of quantum-flips between the four-configurations. 
We show explicitly in Appendix \ref{rotation} that the operator $\tilde{\textbf{L}}_{z,i}$ is the generator of rotations in the space of the four one-particle eigenstates of the operator ${\bf \tilde{L}}_i$. In other words
the operator that rotates the states is
\bea
U = \exp\left({-\imath {\pi\over 2} \tilde{\textbf{L}}_{z,i}}\right) = {\bf 1} - \imath  \tilde{\textbf{L}}_{z,i} -  \tilde{\textbf{L}}_{z,i}^2. 
\eea
\noindent 
The operators $U$ causes transition between
$|\theta\rangle$ and $|\theta-\pi/2\rangle$~:
\begin{equation}\label{U_rot}
{\bf U}_i\left|\theta_i\right\rangle=\left|(\theta-\pi/2)_i\right\rangle \,.
\end{equation}

\subsection{Continuum Model}

The functioning of the rotation operator is more familiar in the continuum model, where the four states are replaced by a continuum of angles $\theta_i$. In fact,
 in the fluctuation regime, the discreteness of the $\theta_i$ variables is a (marginally) irrelevant perturbation and a
continuous distribution of $\theta_i$ gives the correct correlation functions. The model is then just the quantum rotor model :

\bea
H_c = \sum_i \frac{|L_{zi}|^2}{2I} + J\sum_{ij} \cos(\theta_i-\theta_j)
\eea
where $L_{zi} = i\partial /\partial \theta_i$, conjugate to the operator $\theta_i$, causes rotations of $\theta_i$.
The quantum critical fluctuations are calculated by  supplementing the quantum rotor model by the dissipation term of the Caldeira-Leggett form. 
$\exp(i\theta({\bf r},t))$ The Fourier transform of the spectral function of the correlation function $\left<\exp(i\theta({\bf r},t))\exp(i\theta({\bf r}',t'))\right>$ derived in  
\cite{aji-cmv} is

\begin{eqnarray}
\label{eq:flucspec}
\Im\chi({\bf q},\omega) &=& \begin{cases}
 -\chi_0 \tanh(\omega/2T), &|\omega| \lesssim \omega_c;  \\
0,  &|\omega| \gtrsim \omega_c.
\end{cases}
\end{eqnarray}
The value of the cut-off $\omega_c$ and of the amplitude $\chi_0$ will be deduced from experiments below. As noted this is of the same form as suggested  phenomenologically. 
 The spectral weight $\chi_0$ in that equation may be fixed from $\sum_{\bf q}\int d\omega [-\Im\chi({\bf q}, \omega)] \approx (2\Phi_0)^2$, where $\Phi_0$ is the ordered flux in each of the two Cu-O-O
triangular plaquettes in each unit-cell. $\Phi_0$
is given from experiments \cite{fauque, greven} of an ordered moment of about $0.1\mu_B$ per triangular plaquette. In appendix \ref{ap:corr}, we show that the singular part of the correlations of the "angular-momentum" operator, ${\bf L}_{zi}$ are proportional to those in Eq.(\ref{eq:flucspec}).

\section{Coupling of Fluctuations to Fermions}
\subsection{Coupling in the Continuum Model}

It is instructive to write down the coupling for the special case that the conduction electrons are considered well approximated by those in the continuum. This can be done by symmetry considerations alone; the microscopic results for the lattice are shown below to reduce to these in the continuum limit. 

The minimal coupling of the operator for the angular momentum of the collective fluctuations ${\bf L}_{zi}$ to fermions can only be to the local angular momentum of continuum fermions. Thus in the continuum limit, the coupling Hamiltonian of the fluctuations to the fermions is
\bea
\label{coup-cont1}
H_{coup} \propto \int d{\bf r} f(|{\bf r}|) \psi^+({\bf r}) ({\bf r} \times {\bf p}) \psi({\bf r}) {\bf L_z}({\bf r}) + h.c.
\eea
where ${\bf p}$ is the momentum operator, so that $({\bf r} \times {\bf p})$ is the angular momentum operator.
$f(|{\bf r}|)$ is a function which restricts the integrals to be only over the (circular) Wigner-seitz cell of the continuum problem. 
Fourier transforming, we get that 
\bea
\label{coup-cont2}
H_{coup} &\propto& \gamma({\bf k}, {\bf k}')\psi^+({\bf k})\psi({\bf k}') {\bf L_z}({\bf k-k'}) + H.C. \\ \nonumber
\gamma({\bf k}, {\bf k}') &\propto& i ({\bf k} \times {\bf k'})
\eea

The important point about Eq. (\ref{coup-cont2}) is that scattering of fermions through an angle near $\pi/2$ or $-\pi/2$ is strongly favored compared to backward $\pi$ or forward $0$ angles. The other important point is the factor of $i$ signifying coupling to time-reversal breaking fluctuations. These two points are crucial to the the pairing symmetry favored, as shown below. 
It should also be clear that what has been derived is effectively the equivalent for coupling of collective modes which transform as orbital magnetic moments to the orbital moment of fermions, to the familiar coupling $J\psi\left({\bf r}\right) \sigma \psi({\bf r}) \cdot {\bf S}({\bf r})$ of coupling between collective spin-moment variables ${\bf S}$ to the fermion spins. Note that the physics of d-wave being favored through strong enough Antiferromagnetic fluctuations is also related to the scattering through an angle near $\pi/2$ or $-\pi/2$ and a minus sign ($i^2$) due to spin-trace of the fluctuations of ${\bf S}$.

\subsection{Coupling of fluctuations to Fermions for the lattice model} 

Now we return to the lattice model of Eq.(\ref{eqn:mfham}) and Eq.(\ref{coll-ferm-sep}) to generate the coupling of the collective variables to the fermions in microscopic theory. Putting the latter into the last term of the former generates the requisite coupling term:  
\begin{equation}
\label{hcoup1}
H_{coup} = \sum_{i} {V_{pd}\over 16}{\bf L}_{z, i,f}{\bf L^+}_{z,i} + h.c.
\end{equation}
We can now Fourier transform to get the coupling Hamiltonian in momentum space for states in the conduction band. 
Let us take the simplest representation of the conduction band states (in the absence of orbital order) for which the annihilation creation operators are 
\begin{equation}
c_{{\bf k}, \sigma} = \frac{1}{\sqrt{2}}\big(d_{{\bf k}, \sigma} + i(s_x({\bf{k}}) p_{{kx}, \sigma} + s_y({\bf{k}}) p_{{ky}, \sigma} )s_{xy}^{-1}\big).
\end{equation}
Here $s_x({\bf{k}})\equiv \sin(k_xa/2),  s_y({\bf{k}})\equiv \sin(k_ya/2), s_{xy}(k) \equiv \sqrt{s_x^2({\bf{k}})+s_y^2({\bf{k}}))}$.
 Projecting to these states, we have $d_{k}^{\dag}\approx c_{k}^{\dag}/\sqrt{2}$, $p_{kx}^{\dag}\approx -\imath(s_x({\bf k})/\sqrt{2}s_{xy})c_{k}^{\dag}$ and $p_{ky}^{\dag}\approx -\imath (s_y({\bf k})/\sqrt{2}s_{xy})c_{k}^{\dag}$, where $s_{xy} = \sqrt{\sin(k_{x}/2)^{2}+\sin(k_{y}/2)^{2}}$.
The coupling Hamiltonian is
\bea
\label{el-boson-int-1}
H_{coup}&=&\sum_{{\bf k,k'} \sigma} \gamma({\bf k}, {\bf k}') c_{\sigma}^{\dagger}({\bf k}') c_{\sigma}({\bf k})\imath{\bf L}_{z,\textbf{q}} ,
\eea
where the coupling matrix is
\begin{equation}
\gamma({\bf k}, {\bf k}')=\imath
\left({V_{pd}\over {32}}\right)\left[s_x(k)s_y(k') -  s_y(k)s_x(k')\right]S_{xy}(k,k')
\end{equation}
Here  $\textbf{q} = \textbf{k} - \textbf{k}'$ and  $S_{xy}(k,k') =(s_{xy}^{-1}(k)+s_{xy}^{-1}(k'))$. 

\section{Pairing symmetry:}
Integrating over the fluctuations in Eq.(\ref{el-boson-int-1}) in the standard manner gives an effective vertex for scattering of fermion-pairs:
\begin{align}
\label{hpair}
H_{pairing} \approx & \sum_{{\bf k}\sigma{\bf k'}\sigma'} \Lambda({\bf k},{\bf k}')
c^{\dagger}_{\sigma'}(-{\bf k}')c^{\dagger}_{\sigma}({\bf k}')c_{\sigma}({\bf k})c_{\sigma'}(-{\bf k}); \notag \\
\Lambda({\bf k},{\bf k}') = & \gamma(k,k')\gamma(-k,-k') \Re\chi(\omega=\epsilon_{\bf k}-\epsilon_{\bf k}').
\end{align}
The susceptibility appearing in the coupling $\Lambda({\bf k},{\bf k}')$ is given in eq.\ref{eq:flucspec}. For further discussion see appendix \ref{ap:corr}.  As will be discussed, this is correct to $O(\lambda {\omega_c}/{E_f})$, where $\lambda$'s are dimensionless coupling constants derived below. 

It is illuminating to note first the symmetry of the favored pairing due to such a coupling in a continuum approximation for fermions near the fermi-energy.  In this approximation, $s_x({\bf k})
\propto(k_xa)/2$, etc. so that $\gamma({\bf k}, {\bf k}') \propto  i({\bf k} \times {\bf k}')$. The pairing vertex is then
\begin{equation} \label{kxk'}
\Lambda\left(\textbf{k},\textbf{k},\right) \propto - ({\bf k} \times {\bf k}') ^2\Re \chi(({\bf k}-{\bf k'}),\omega).
\end{equation}
Since  $\Re \chi({\bf k}-{\bf k'},\omega) < 0$  for $-\omega_c <\omega < \omega_c$, independent of momentum, the pairing symmetry is given simply by expressing $({\bf k} \times {\bf
k}') ^2$ in separable form~:
\begin{align} ({\bf k} \times {\bf k}') ^2 &= 1/2 \left[(k_x^2+k_y^2)(k_x^{'2}+k_y^{'2})
-  (k_x^2-k_y^2)(k_x^{'2}-k_y^{'2})\right.\nonumber \\
&- \left. 4(k_xk_y) (k'_x k'_y)\right].
\end{align}
Pairing interaction in  the $s$-wave channel is repulsive, that in the two $d$-wave channels is equally attractive, and in the odd-parity channels is zero. The factor $i$ in $\gamma({\bf k}, {\bf k}')$, present because the coupling is to fluctuations of 
time-reversal odd operators, is crucial in determining the sign of the interactions of the pairing vertex.

We now return to Eq.(\ref{hpair}) for a semi-quantitative analysis of pairing in the lattice. The dimensionless coupling constants $\lambda_{\alpha}$ determining $T_c$ and the gap $\Delta$ are given in a generalized Eliashberg theory by \cite{msv}
\begin{equation}
\lambda_{\alpha} = 2N(0) \int{d}\Omega{d}\Omega' \Lambda({\bf k},{\bf k}')F_{\alpha}({\bf
k})F_{\alpha}({\bf k}')
\end{equation}
Here $d\Omega=N(0)^{-1}dS_k/|v_k|$ where $S_k$ is an element the Fermi surface and $N(0)$ is an effective density of states on the Fermi surface for one spin species. Also,  $F_{\alpha}({\bf k})F_{\alpha}({\bf k}')$ are separable lattice harmonics on to which $\Lambda({\bf k},{\bf k}')$ is projected \cite{msv}:
\begin{align}\label{channels}
F_{s}({\bf k}) =&  N_{s}[1-\cos(k_xa)\cos(k_ya)];  \notag\\
F_{d1}({\bf k}) =& N_{d1}[\cos(k_xa) -
\cos(k_ya)];  \notag\\
F_{d2}({\bf k})  =& N_{d2} \sin(k_xa)\sin(k_ya)\,.
\end{align}
Where the labels $(s,d1,d2)$ represent the irreducible lattice representations $(A_{1g},B_{1g},B_{2g})$ of a tetragonal lattice, popularly referred to as extended s wave, $d_{x^{2}-y^{2}}$ and $d_{xy}$ symmetries, respectively. The factor $N_{\alpha}$ ensures  normalization $\int{d}\Omega
F_{\alpha}(k)F_{\alpha}(k) = 1$.

The resolution of $\Lambda(k,k')$ in Eq.\ref{hpair}
is
\begin{equation}
\label{Lambda}
\Lambda(k,k') = \lambda_0\Big[ \frac{F_{s}({\bf k})F_{s}({\bf k'})}{N_{s}^2} -\frac{F_{d1}({\bf k})F_{d1}({\bf k'})}{N_{d1}^2} - \frac{F_{d2)}({\bf k})F_{d2}({\bf k'})}{N_{d2}^2} \Big]
\end{equation}

From (\ref{Lambda}), the $s$-wave interactions are repulsive while the interaction is equally attractive for the $d(x^2-y^2)$ and $d_{xy}$-waves for a circular
fermi-surface. For the actual fermi-surface of the cuprates in which the fermi-velocity is largest in the $(1,1)$ directions and the least in the $(1,0)$ or the Cu-O bond-directions,
$d(x^2-y^2)$-pairing is favored because in that case the maximum gap is in directions where the density of states is largest.

\section{Deduction of Parameters from ARPES experiments and Estimates of $T_c$ and $\Delta$}

In this section, we first summarize experimental evidence and calculation directly showing that the scattering of fermions is uniquely given by the derived QCF's and how ARPES experiments have been used to determine the parameters used later in this paper. We argue that since superconductivity is an instability of the normal state which occurs at $T < T_c$, it is unlikely that any other fluctuations can dominate in determining $T_c$. 

The value of the cut-off $\omega_c$, of the amplitude $\chi_0$ in Eq. \ref{eq:flucspec} and the coupling constants in front of $\gamma$ in Eq. \ref{hpair} can be provided in terms of the parameters of the microscopic model as well as deduced from experiments below. Given
a coupling function of such fluctuations to fermions $\gamma({\bf k},{\bf k'})$ to scatter from ${\bf k}$ to ${\bf k}'$, calculated below, the self-energy of the fermions is \cite{zhu/waterfall}  
\begin{eqnarray}
\label{selfenergy}
\Im\Sigma(\omega, {\bf k}) &=& -\frac{\pi}{2} \lambda({\bf k})\begin{cases}
 |\omega|, & |\omega| \lesssim \omega_c  \\
 \omega_c, & |\omega| \gtrsim \omega_c.
\end{cases}
\end{eqnarray}
Here $\lambda({\bf k}) = N(0)<\gamma^2>_{k'}$ and $<\gamma^2>_{k'}$ is the average of $|\gamma({\bf k},{\bf k'})|^2$ over ${\bf k}'$ on the fermi-surface. In the phenomenological approach \cite{mfl} this was taken to be momentum independent. 
\begin{figure}[ht!]
  \centerline{
  \includegraphics[width=0.8\columnwidth]{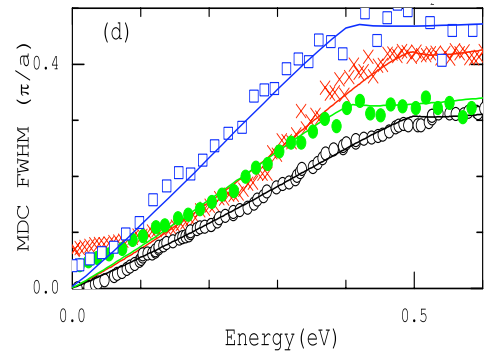}}
  \caption{The MDC linewidths along the $\pi,pi$ directions for all the measured cuprates. The detailed references for each cuprate are given in Ref.(\onlinecite{zhu/waterfall})}
   \label{mdcwidth}
\end{figure}
In Ref.(\onlinecite{zhu/waterfall}), this expression is compared with the data in all available directions and parts of the phase diagram of the cuprates. In fig.(\ref{mdcwidth}), we show the deduced MDC linewidth in the $(\pi,\pi)$ direction for all the cuprates near optimal doping for which data is available. This data is taken with poor energy resolution, $ \gtrsim 40meV$, to cover a wide energy range. The linearity of the linewidth with energy in the normal state for low energies has been checked with better precision in other experiments. Here we focus on the full energy range. We notice the remarkable correspondence with Eq.(\ref{selfenergy}) with the cut-off $\omega_c$ between 0.4 and 0.5 eV for all the measured cuprates. Below we will use the slope of these curves for $\omega \lesssim \omega_c$ to deduce the coupling constant $\lambda_{\ell}$ in different angular momentum channels. The normal state resistivity and optical conductivity can also be calculated using the values of $\lambda_{\ell}$ and $\omega_c$ to within about $30\%$ of those deduced from the single-particle spectra.

The most important point to be noted from fig. (\ref{mdcwidth}) is the following:  The result in Eq. (\ref{selfenergy}) arises because the scattering at any energy $\omega >> T$ is proportional to the integrated weight of fluctuations up to $\omega$, i.e  $\propto \int_0^\omega \Im \chi(\omega')$. Therefore the linearity of the scattering rate with $\omega$ up to about  $\omega_c$ and constancy thereafter is a direct proof of the fluctuations of the form of Eq. (\ref{eq:flucspec}). The rather sharp $\omega_c$ proves that one need not be concerned that a distinct energy scale of fluctuations may not exist \cite{pwa}.

An important deduction from recent analysis \cite{bok}  of high resolution laser based ARPES at different angles on the fermi-surface is that a momentum independent fluctuation spectra is obtained from the inversion of the data through Eliashberg equations to fit the data at different angles. 

In Eq.(\ref{Lambda}), $\lambda_0$ may be estimated thus: $\Re\chi(\omega) \approx -2\chi_0 \ln \frac{\omega_c}{|\omega|}$ for $|\omega|\lesssim \omega_c$ and it is vanishing beyond. The cut-off $\omega_c$ is
important but the weak dependence on $|\omega|$ may be ignored by replacing it by $O(T_c)$ for estimates of parameters determining $T_c$. For $\omega_c \approx 0.4eV$, as deduced from
experiments \cite{zhu/waterfall} and $T_c \approx 100 K$, $\Re\chi(\omega)$ is then $\approx -6\chi_0$ for $\omega \lesssim \omega_c$. Using this estimate $\lambda_0 \approx -6V^2N(0) \chi_0$
with a cut-off in the range of interaction at $\omega_c$. (Here $s_{xy}^2(k=k_F) \approx 1/2$ has been used.)

We now ask whether the $\omega_c$ and $\lambda$
deduced from experiments in the normal state yield the right order of magnitude of both $T_c$'s and
the ratio of the zero-temperature gap $\Delta$ to $T_c$.  Compared to $s$-wave
superconductors, the normal self-energy and inelastic scattering
lead to a stronger depression in $T_c$ and a stronger enhancement in
$\Delta/T_c$ in $d$-wave superconductors \cite{msv};  
estimate, $T_c \approx \omega_c\exp(-(1+|\lambda_s|)/|\lambda_d|)$, where $\lambda_s$ is the
coupling constant which appears in the normal self-energy and $\lambda_d$, the coupling constant which appears in the anomalous self-energy.
From
Eq.~(\ref{channels}), $|\lambda_d|/|\lambda_s| \approx 1/2$. The
normal state self-energy deduced from experiments
\cite{zhu/waterfall} gives $|\lambda_s| \approx 1$ and $\omega_c
\approx 0.4 eV$. The formula above
gives $T_c\approx 80K$ for these values.

The value of $\Delta/T_c$ may be read off from Fig.~(4) of Ref.~(\onlinecite{msv}) using the parameters above to be about $2.5\Delta_0/T_{c0} $, where $\Delta_0$ and $T_{c0}$ are estimated
ignoring the self-energy and inelastic scattering. This gives $\Delta/T_c \approx 5$. $\Delta/T_c$ of 4-5 are reported for the cuprates \cite{pashupathi}.

We can roughly estimate the value of $\lambda$'s and $\omega_c$ from the microscopic parameters. As estimated above, $|\lambda_s| \approx
6N(0)V^2\chi_0$. To get $\lambda_s \approx 1$ requires $\chi_0 \approx 6\times 10^{-2} (eV)^{-1}$ for $N(0)\approx 2$ states$/eV/unit-cell, V\approx 1 eV$. For the deduced $\omega_c \approx
0.4 eV$, this requires an ordered $\Phi_0$ per triangular plaquettes of about $O(0.1)$. As mentioned already this corresponds well with the measured moment deduced by Fauque et al.
\cite{fauque}.

A comment should be made on the validity of a theory of superconductivity
with electron-electron interactions using an Eliashberg type
simplification \cite{pwa}. The Eliashberg simplification of the theory
works only if there is a small parameter which limits the irreducible vertex in the particle-particle channel to
ladder diagrams alone with the simplest single-particle self-energy in the propagators. For any model with single-particle self-energy which is nearly momentum
independent in the normal state as in the present case (and the case of electron-phonon interactions), this parameter is $O(\lambda \omega_c/W)$, where
$W$ is the {\it bare} electronic band-with. This parameter for the
cuprates with $W \approx 2 eV$ is $O(1/5)$. This does not allow the luxury of a parameter of $O(10^{-2})$ as
for electron-phonon interactions but small enough to have a systematic theory.

The results of this paper and their applicability for a consistent calculation of superconducting parameters, can be tested in detail through inversion of Angle Resolved Photoemission (ARPES)
data by a procedure, which is a generalization of the Rowell-MacMillan procedure for $s$-wave superconductors \cite{vekhter}. (Approximate inversion of such data has recently appeared \cite{carbotte} with results consistent with the results of this paper for the shape of the spectrum, and its cut-off energy.)

\section{Concluding Remarks}

This paper has derived that the fluctuations responsible for the normal state anomalies near the QCP couple to fermions to promote d-wave pairing and with parameters taken from fits to the normal state data and consistent with estimates from microscopic calculations calculated the right magnitude of both $T_c$ and $\Delta/T_c$. The critical fluctuation spectra develops a low-energy cutoff which increases as $x$ increases. So $T_c$ is expected to fall. For underdoping, superconductivity can only be calculated in a state with the competing order parameter whose strength goes up as $x$ decreases from $x_c$. Again $T_c$ must fall. The details of such calculations are work for the future.

Useful discussions with Han-Young Choi and Lijun Zhu are gratefully acknowledged. CMV's research was partially supported by National Science Foundation grant DMR- 
0906530 \\

\appendix
\section{Introduction to the Appendices}\label{Intro}
There are three principal purposes of these appendices. First is to derive the Hamiltonian, Eq. (\ref{eqn:mfham}) from the interactions of Eq.(\ref{nn-int}). Towards this end, we first give the expressions for the Kinetic energy so as to specify the choice of phases made in the $d$ and the $p$-orbitals and of the current operators in the gauge chosen for the calculations. This is followed by a detailed derivation of Eq.(\ref{eqn:mfham}). The second purpose is to show that the operator ${\bf L}_{iz}$ provides the kinetic energy which acts as a rotation operator for the four possible flux patterns given by ${\bf L}_{i\ell}$ in the cell $i$. The third is to show that the correlations of the operators  ${\bf L}_{iz}$ have the same singularities as the order parameter fluctuations calculated in Refs. (\onlinecite{aji-cmv, aji-cmv/warps}).

\section{ Kinetic Energy} \label{KE}
\begin{widetext}
We use a choice of the phase of the $d$-orbitals so that the starting kinetic energy in the model is a real operator. With this choice and the notation specified in Fig.(\ref{fig:unitcell}), the kinetic energy operator is
\begin{eqnarray}\label{ke}
K.E. &=&   \sum_{i}\left[ {t_{pd}\over 2}\left(d_{i1}^{\dagger}p_{i1x} +  d_{i4}^{\dagger}p_{i4x} - d_{i2}^{\dagger}p_{i1x} -d_{i3}^{\dagger}p_{i4x} + d_{i1}^{\dagger}p_{i1y}  +d_{i2}^{\dagger}p_{i2y} - d_{i4}^{\dagger}p_{i1y} - d_{i3}^{\dagger}p_{i2y}  \right)\right.\\ \nonumber
&+&\left. {t_{pp}}\left(p_{i1x}^{\dagger}p_{i2y}-p_{i2y}^{\dagger}p_{i4x}+p_{i4x}^{\dagger}p_{i1y}-p_{i1y}^{\dagger}p_{i1x}\right)+h.c\right]\\ \nonumber
\end{eqnarray}
\end{widetext}

\section{Gauge invariant combination of the vector potentials}\label{fluxes}

The complex hopping matrix elements on the links in the unit-cell are equivalent to vector potentials living on the links of the unit-cell. Flux operators are formed by sum of the phase difference (or vector-potentials) in closed loops of links which form independent areas in each unit-cell.
\begin{figure}
  \includegraphics[width= 0.6\columnwidth]{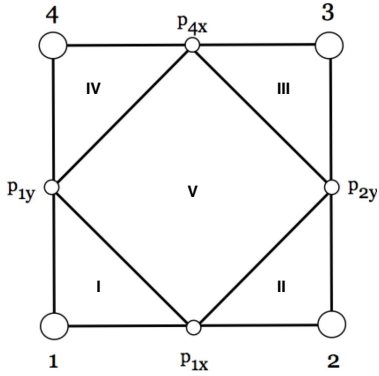}
 \caption{ The five areas in which a unit-cell is divided by connecting Cu-O and O-O links. Sum of directed Current operators on the links to form closed loops define five flux-operators in the areas marked by Roman letters. These can be further combined to form 5 flux patterns with the point group symmetry of the square lattice as in fig.(\ref{fig:fluxes}).}     
 \label{fig:5fluxes}
\end{figure}

There are 12 Cu-O and O-O links shown in Fig.(\ref{fig:unitcell}) but 8 of these are shared by the adjoining unit-cells, so that there are only 8 links per unit-cell. But the links form only five areas per unit-cells as shown in Fig. (\ref{fig:5fluxes}). This is consistent with the fact that there are three sites per unit-cell in which independent gauge transformation can be made to obtain 5 gauge invariant flux-operators per unit-cell. 
\begin{figure}
  \includegraphics[width=\columnwidth]{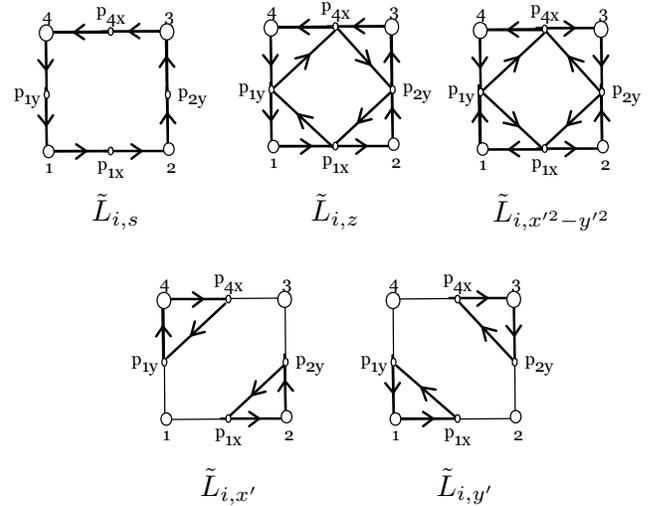}
 \caption{ The five gauge invariant combinations of link variables that respect the symmetry of the lattice. $\tilde{{\bf L}}_{i,1}$ has the symmetry of an overall flux in an unit cell. $\tilde{{\bf L}}_{i,2}$ has zero total flux in an unit cell. The corresponding operator has the symmetry of an angular momentum. Uniform ordering of $\tilde{{\bf L}}_{i,3}$ results in the $\Theta_{I}$ phase. $\tilde{{\bf L}}_{i,4}$ and $\tilde{{\bf L}}_{i,5}$ are the operators that condense to give the $\Theta_{II}$ phase.}
     \label{fig:fluxes}
\end{figure}

The  five closed loops in any unit-cell can be combined to form five new combinations which with the point-group symmetries of the lattice.  The five combinations are pictorially represented in fig.\ref{fig:fluxes}.
The five combinations of {\it flux}-operators $\tilde{{\bf L}}_{i,\mu}$ are,
\begin{widetext}
\begin{eqnarray}\label{mfdecoupling}
\tilde{{\bf L}}_{i,s} &=& {\imath}\left(d_{i1}^{\dagger}p_{i1x} -  d_{i4}^{\dagger}p_{i4x} + d_{i2}^{\dagger}p_{i1x} -d_{i3}^{\dagger}p_{i4x} - d_{i1}^{\dagger}p_{i1y}  +d_{i2}^{\dagger}p_{i2y} - d_{i4}^{\dagger}p_{i1y} + d_{i3}^{\dagger}p_{i2y}  \right)+h.c\\ \nonumber
\tilde{{\bf L}}_{i,\overline{s}} &=& {\imath}\left(d_{i1}^{\dagger}p_{i1x} -  d_{i4}^{\dagger}p_{i4x} + d_{i2}^{\dagger}p_{i1x} -d_{i3}^{\dagger}p_{i4x} - d_{i1}^{\dagger}p_{i1y}  +d_{i2}^{\dagger}p_{i2y} - d_{i4}^{\dagger}p_{i1y} + d_{i3}^{\dagger}p_{i2y}  \right)+h.c\\ \nonumber
&-& {\imath}\left(p_{1x}^{\dagger}p_{2y}- p_{2y}^{\dagger}p_{4x}+p_{4x}^{\dagger}p_{1y}-p_{1y}^{\dagger}p_{1x}\right)+h.c\\ \nonumber
\tilde{{\bf L}}_{i,x'^{2}-y'^{2} }&=& {\imath}\left(-d_{i1}^{\dagger}p_{i1x} - d_{i4}^{\dagger}p_{i4x} +d_{i2}^{\dagger}p_{i1x} +d_{i3}^{\dagger}p_{i4x} +d_{i1}^{\dagger}p_{i1y}  +d_{i2}^{\dagger}p_{i2y} -d_{i4}^{\dagger}p_{i1y} -d_{i3}^{\dagger}p_{i2y}  \right)+h.c\\ \nonumber
&-& {\imath}\left(p_{1x}^{\dagger}p_{2y}+ p_{2y}^{\dagger}p_{4x}+p_{4x}^{\dagger}p_{1y}+p_{1y}^{\dagger}p_{1x}\right)+h.c\\ \nonumber
\tilde{{\bf L}}_{i,x'} &=& {\imath}\left( d_{i2}^{\dagger}p_{i1x} +d_{i4}^{\dagger}p_{i4x} + d_{i4}^{\dagger}p_{i1y} + d_{i2}^{\dagger}p_{i2y}  \right)+c.c\\ \nonumber
&-& {\imath}\left(p_{1x}^{\dagger}p_{2y}-p_{4x}^{\dagger}p_{1y}\right)+h.c\\ \nonumber
\tilde{{ \bf L}}_{i,y'} &=& {\imath}\left(d_{i1}^{\dagger}p_{i1x} +  d_{i4}^{\dagger}p_{i4x}  - d_{i1}^{\dagger}p_{i1y} - d_{i3}^{\dagger}p_{i2y}  \right) + h.c\\ \nonumber
&-&{\imath}\left(p_{2y}^{\dagger}p_{4x}-p_{1y}^{\dagger}p_{1x}\right)+ h.c\\ \nonumber
\end{eqnarray}
\end{widetext}

As will be evident from Fig.(\ref{fig:fluxes}), $\hat{\textbf{L}}_{s}$ has the identity representation of the flux-operator, while $\hat{\textbf{L}}_{\overline{s}}$ in common parlance may be called the "extended s-wave" representation. $\hat{\textbf{L}}_{x'}$ and $\hat{\textbf{L}}_{y'}$ are the operators that have the symmetry of the $\Theta_{II}$ phase \cite{cmv}. The phases produced by their condensation have {\it magneto-electric} symmetry, describable by time-reversal odd polar-vectors pointing in the $\hat{{\bf x'}}, \hat{\bf{y'}}=(\hat{{\bf x}}\pm \hat{\bf y})/\sqrt{2}$ directions respectively. $\hat{\textbf{L}}_{x'^{2}-y'^{2}}$ has the symmetry of the $\Theta_I$ phase described earlier \cite{cmv}.

\section{ Derivation of the Interaction Hamiltonian in terms of flux operators}\label{loopham}
In this appendix, we give details of the re-expression of the Cu-O and O-O interaction Hamiltonian $H_{nn}$ given by 
Eq.(\ref{nn-int}) in terms of current operators using the operator identity of Eq. (\ref{opident}). The purpose of doing this is to derive the relevant part of the interaction Hamiltonian $H_{nn}$ in terms of the irreducible combinations of the {\it flux-operators} of Eq.(\ref{irred-rep-flux}). 
The unimportant one-electron terms in Eq.(\ref{opident}) are ignored and only the spin-diagonal parts are kept. The spin-diagonal part of the interaction $H_{nn}$ across the 12 links in Fig.(\ref{fig:unitcell}) 
\begin{eqnarray}
\label{j'ells-n}
H_{nn} &=& - \frac{V_{pd}}{4}\sum_{\ell=1}^{4}\left[ |{\bf \mathcal{O}}_{i,\ell, x}|^2+ |{\bf \mathcal{O}}_{i,\ell, y}|^2 \right ] \\ \nonumber
&- &\frac{V_{pp}}{2}\sum_{\ell=1}^{4}|{\bf \mathcal{O}}_{i,\ell, xy}|^2.
\end{eqnarray}
Here $\ell$ sums the four Cu-sites per unit-cell and the operators ${\bf \mathcal{O}}$, given after Eq. (\ref{opident}) are written down again here:
\begin{eqnarray}
\label{defs-n}
{\bf \mathcal{O}}_{i,\ell,x} &=& \sum_{\sigma} i d_{i, \ell, \sigma}^+ p_{i, \ell, x, \sigma} + h.c., ~etc, \\
{\bf \mathcal{O}}_{i,\ell,x,y} &=& \sum_{\sigma} i  p_{i, \ell, y, \sigma}^+ p_{i, \ell, x, \sigma} + h.c.,~ etc. 
\end{eqnarray}

\begin{figure}[ht!]
  \centerline{
  \includegraphics[width=0.6\columnwidth]{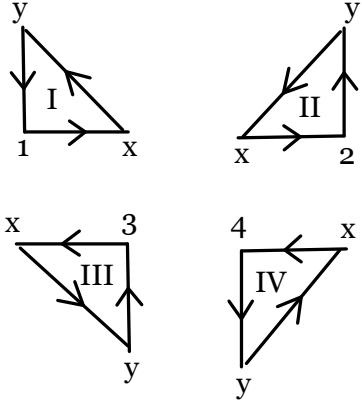}}
  \caption{Convention used for defining the triangles and the link operators.}
   \label{triangles_def}
\end{figure}

\noindent To simplify notation, we use a slightly modified labeling scheme in this appendix. In a given unit cell there are four triangles with one of their vertices being a  Cu site and one square with its vertices being the four Oxygen atoms.
The subscript $x$ and $y$ refer to the $p_{x}$ and $p_{y}$ orbital, that combined with the Copper site, labeled by $\ell$, form the triangle. Explicitly the triangles are $\{1,p_{1x},p_{1y}\}$, $\{2,p_{1x},p_{2y}\}$, $\{3,p_{4x},p_{2y}\}$ and $\{4,p_{4x},p_{1y}\}$. The flux in the triangles labelled $\textrm{L} = I, ...IV$ in Fig.(\ref{fig:5fluxes})  with a Cu-site at $\ell = 1,..4$ is, 
\begin{eqnarray}
\label{flux-I}
f_{i, I} &\equiv & {\bf \mathcal{O}}_{i,1,x} - {\bf \mathcal{O}}_{i,1,y} + {\bf \mathcal{O}}_{i,1,xy}.\\ \nonumber
f_{i, II} &\equiv &{\bf \mathcal{O}}_{i,2,x} + {\bf \mathcal{O}}_{i,2,y} + {\bf \mathcal{O}}_{i,2,xy}.\\ \nonumber
f_{i, III} &\equiv &{-\bf \mathcal{O}}_{i,3,x} + {\bf \mathcal{O}}_{i,3,y} + {\bf \mathcal{O}}_{i,3,xy}.\\ \nonumber
f_{i, IV} &\equiv &{-\bf \mathcal{O}}_{i,4,x} - {\bf \mathcal{O}}_{i,4,y} + {\bf \mathcal{O}}_{i,4,xy}.
\end{eqnarray}
A clockwise choice of currents around the loop is chosen to define $f$ to be positive. The convention is shown in fig.\ref{triangles_def}. 

Similarly, the fifth flux, $f_{i,V}$ operator (See Fig. (\ref{fig:fluxes}))
is
\begin{eqnarray}
\label{flux-v}
f_{i, V} = \sum_{\textrm{L}} f_{i,\textrm{L}}-2 {\bf \mathcal{O}}_{i,\ell,xy}.
\end{eqnarray}
Here the sum of the currents on the Cu-O links cancels the current in the O-O links such that there is no flux in the corner triangles.

\noindent For the triangle $\textrm{L}=1$
\begin{eqnarray}
\label{decomp-1}
{V_{pd} \over 2}\left( |{\bf \mathcal{O}}_{i,1, x}|^2+ |{\bf \mathcal{O}}_{i,1, y}|^2 \right ) &=& \left({V_{pd}\over 4}\right)\left( |({\bf \mathcal{O}}_{i,\ell, x} + {\bf \mathcal{O}}_{i,\ell, y}|^2\right.   \nonumber\\ 
&+& \left. |({\bf \mathcal{O}}_{i,\ell, x} - {\bf \mathcal{O}}_{i,\ell, y}|^2 \right).
\end{eqnarray}
We now note that the first term in Eq.(\ref{decomp-1}) has finite sum of currents at the Cu-sites and can form closed loops only by adding similar terms from neighboring cells. These cannot give rise to either $q=0$ order or long wavelength fluctuations and are not considered further. Now we add  $(V_{pd}/4) |{\bf \mathcal{O}}_{i,\ell,x,y}|^2$ to the second term in Eq.(\ref{decomp-1}) and subtract it from the similar term with coefficient ~$V_{pp}$ in Eq.(\ref{j'ells-n}) so that we can write 
\bea
\label{decomp-2}
H_{nn}^{\textrm{L}=1} &=& -
 \frac{V_{pd}}{8}\left( |({\bf \mathcal{O}}_{i,1, x} - {\bf \mathcal{O}}_{i,1, y} + {\bf \mathcal{O}}_{i,1, x,y})|^2\right. \\ \nonumber 
&+& \left. |({\bf \mathcal{O}}_{i,1, x} - {\bf \mathcal{O}}_{i,1, y} - {\bf \mathcal{O}}_{i,1, x,y})|^2 \right). \\ \nonumber
&-& \left. \frac{(2V_{pp} - V_{pd})}{4} |{\bf \mathcal{O}}_{i,1,xy}|^2 \right).
\end{eqnarray}
Note using Eq.(\ref{flux-I}) 
that the first term in Eq.(\ref{decomp-2}) is equal to $f_{i,\textrm{L}=1}^{2}$. This exercise can be repeated for $\textrm{L}=2,3$ and $4$. For each triangle we get three terms: 1) $f_{i,\textrm{L}}^{2}$, 2) term analogous to the second term in Eq.\ref{flux-I} and 3)  $ ((2 V_{pp} - V_{pd})/4)|{\bf \mathcal{O}}_{i,\ell,xy}|^2 $. We can sum over $\ell$ in the second term to produce one combination which (see Eq.(\ref{flux-v}) is $~ |f_{i, V} |^2$. The other three can be removed by gauge transformations at the 3 sites in each unit-cell. 

 Next note that
\begin{eqnarray}
\tilde{{\bf L}}_{i,x'} &=&   f_{i, I} - f_{i, III}   \\ \nonumber
\tilde{{\bf L}}_{i,y'} &=&  f_{i, II} - f_{i, IV}   \\ \nonumber
\tilde{{\bf L}}_{i,x'^2-y'^2} &=& f_{i, I} + f_{i, III} -f_{i, II} - f_{i, IV}    \\ \nonumber
\tilde{{\bf L}}_{i,\overline{s}} &=&  {1\over 2}( -f_{i, V} + \sum_{L = I, IV} f_{i, L})   \\ \nonumber
\tilde{{\bf L}}_{i,{s}} &=&{1\over 2} \sum_{L = I,V} f_{i, { L}}    
\end{eqnarray}

\noindent We also note that 
\bea
\sum_{\ell=1}^{4}  {\bf \mathcal{O}}_{i,\ell,x,y} = \tilde{{\bf L}}_{i,\overline{s}}
\eea
as well.

Using the above identities, we can write the gauge invariant part of $H_{nn}$ as 
\begin{eqnarray}
\label{decomp-3}
H_{nn}&= & -({V_{pd}\over 16}) \left( |\tilde{{\bf L}}_{i,x'}|^2 +  |\tilde{{\bf L}}_{i,y'}|^2\right. \\ \nonumber
&+ &({1\over 2})|\tilde{{\bf L}}_{i,x'^2-y'^2}|^2 
+\left.  |\tilde{{\bf L}}_{i,s}|^2 \right)\\ \nonumber
&-&\left({{ V_{pp}}\over {8}}\right)  |\tilde{{\bf L}}_{i,\overline{s}}|^2 
\end{eqnarray}
This is in the desired form. 

Let us also define an operator
\bea
\tilde{{\bf L}}_{i,z} \equiv \sum_{L = I,..,IV} f_{i, L}.
\eea
A term 
\bea
H_{ke} = -\frac{V_{pd}}{16} |\tilde{{\bf L}}_{i,z}|^2
\eea
is also present in the interactions. 
We are concerned in this paper  with the fluctuations of the observed phase which is realized by the {\it local} condensation of the collective parts of $\hat{\textbf{L}}_{x'}$ and $\hat{\textbf{L}}_{y'}$ into four possible domains in any unit-cell. In the quantum-fluctuation regime, the important fluctuations are between these four configurations in any unit-cell. As we show below, such fluctuations are caused by the operator $\hat{\textbf{L}}_{z}$ so that $H_{ke}$ acts as the kinetic energy operator rotating the configurations of $\hat{\textbf{L}}_{x'}$ and $\hat{\textbf{L}}_{y'}$. The relevant terms in the above must therefore include $H_{ke}$ beside those involving the  $\hat{\textbf{L}}_{x'},\hat{\textbf{L}}_{y'}$ operators. This completes the derivation of  Eq.(\ref{eqn:mfham}) of Sec. II.

 The operator $\tilde{{\bf L}}_{i,s}$ corresponds as noted to a net flux in the unit-cell. For it to order, its expectation value must reverse between neighboring cells giving rise to flux patterns with broken translational symmetry. Such patterns are not observed in experiments. What about the uniform ordering of $\tilde{{\bf L}}_{i,s}$? This would correspond to a net flux in the sample or macroscopic boundary currents. By a general theorem, such ordering is impossible  because long-wavelength variations of $\tilde{{\bf L}}_{i,s}$ have all the symmetries of magnetic field produced by a vector potential, which cannot acquire mass. However from the same considerations, fluctuations of such operators at long wavelengths have the properties of photons and therefore may be quite important in the pseudogap phase. The propagator of such fluctuations and its coupling to fermions and the consequences of this coupling will be presented in the near future.

\section{Properties of $\tilde{{\bf L}}_{i,z}$}\label{rotation}

In this appendix we show that the operator $\hat{\textbf{L}}_{z}$ defined on the lattice is a generator for rotations. To do so let us consider the four states represented pictorially in fig.\ref{fig:domain} and write down the one particle wavefunction which has the same expectation value of the current operator as that in the collective states ${\bf L}_i$ in a unit cell.  Define a set of basis operators centered at each of the four copper sites in a cell $i$ labelled by their center, $\ket{\psi_{i1}}^{--}$, $\ket{\psi_{i2}}^{--}$, $\ket{\psi_{i3}}^{--}$, $\ket{\psi_{i4}}^{--}$: 

\begin{align}\label{-x-y}
\psi_{i1}^{--} ={1\over 2}(\sqrt{2}d_{i1} + e^{\imath\phi} p_{i1x} - e^{\imath\phi}p_{i1y}) \notag \\
\psi_{i2}^{--} = {1\over 2} (\sqrt{2}d_{i2} + e^{\imath\phi} p_{i1x} - e^{\imath\phi} ip_{i2y})  \notag \\
\psi_{i3}^{--} =  {1\over 2}(\sqrt{2}d_{i3} + e^{\imath\phi} p_{i4x} - e^{\imath\phi} ip_{i2y})  \notag \\
\psi_{i4}^{--} = {1\over 2} (\sqrt{2}d_{i4} + e^{\imath\phi} p_{i4x} - e^{\imath\phi} ip_{i1y})  \notag \\
\end{align}

\noindent In terms of these operators, the wavefunction of the electron corresponding to, say, the state $\left|\theta = 225^{0}\right>$ $=$ $\left|-\hat{x}-\hat{y}\right>$ of ${\bf L}_i$ is

\begin{equation}\label{wavefunction-x-y}
\left|-\hat{x}-\hat{y}\right> ={1\over 4}\left( \psi_1^{--} +\psi_2^{--}+ \psi_3^{--}+ \psi_4^{--} \right)^{\dag}\left|0\right>
\end{equation}
Similar representation of the other four-states in terms of fermion operators may be written down. The state $\left|\theta = 135^{0}\right>$ $=$ $\left|-\hat{x}+\hat{y}\right>$ is obtained by rotating clockwise by $90^{0}$. The corresponding rotation operator is $U\left(-\pi/2\right) = e^{-\imath \hat{\textbf{L}}_{z} \left(\pi/2\right)}$ is

\begin{equation}
U\left(-\pi/2\right) = e^{-\imath \hat{\textbf{L}}_{z} \left(\pi/2\right)} = \textbf{I} - \imath  \hat{\textbf{L}}_{z} -  \hat{\textbf{L}}_{z} ^{2}
\end{equation}

\noindent Using 

\bea
[d_2^+d_3 + d_3^+d_1, j_{12}] =  j_{31},
\eea

\noindent  and  $[j_{12},j_{23}] =ij_{13}$ we can explicitly verify that

\begin{eqnarray}
\left[ U\left(-\pi/2\right), {1\over 4}\left( \psi_1^{--} +\psi_2^{--}+ \psi_3^{--}+ \psi_4^{--} \right)^{\dag}\right] \\ \nonumber
~~~~= {1\over 4}\left( \psi_1^{-+} +\psi_2^{-+}+ \psi_3^{-+}+ \psi_4^{-+} \right)^{\dag}
\end{eqnarray}

\noindent where
\begin{align}\label{-x+y}
\psi_{i1}^{-+} ={1\over 2}(\sqrt{2}d_{i1} + e^{\imath\phi} p_{i1x} +e^{\imath\phi}p_{i1y}) \notag \\
\psi_{i2}^{-+} = {1\over 2} (\sqrt{2}d_{i2} + e^{\imath\phi} p_{i1x} + e^{\imath\phi} ip_{i2y})  \notag \\
\psi_{i3}^{-+} =  {1\over 2}(\sqrt{2}d_{i3} + e^{\imath\phi} p_{i4x} + e^{\imath\phi} ip_{i2y})  \notag \\
\psi_{i4}^{-+} = {1\over 2} (\sqrt{2}d_{i4} + e^{\imath\phi} p_{i4x} + e^{\imath\phi} ip_{i1y}).  \notag \\
\end{align}

\noindent The exercise can be repeated to show that the operator indeed rotates the states of $\hat{\textbf{L}}$ clockwise by $90^{0}$. The lattice fermion operator $\hat{\textbf{L}}_{z}$ is the generator of rotation and corresponds to the angular momentum operator in the continuum limit.

\section{ Correlations of the angular momentum operator}\label{ap:corr}

The coupling of the fermions to the fluctuations of the order parameter are proportional to the angular momentum operator ${\bf L}_{z} $. The pairing interaction that leads to superconducting instability involve the  $\left<{\bf L}_{z i} {\bf L}_{z j}\right>$ correlation function. To analyze the connection with the $\left<\exp(\imath\theta({\bf r},t))\exp(\imath\theta({\bf r}',t'))\right>$ correlation derived in Aji and Varma\cite{aji-cmv} we first must identify the appropriate representation of the field $\exp\left(\imath \theta({\bf r}',t')\right)$.  In the path integral formulation the correlation function computed is

\begin{equation}
\label{orderparameter} C_{i,j,\tau,\tau '} = \left<e^{\imath
\theta_{i\tau}}e^{-\imath\theta_{j\tau '}}\right> 
\end{equation}

\noindent where $\theta_{i\mu}$ is the classical phase at site $i$ at time $\mu$. In this calculation the standard procedure of slicing time into infinitesimal segments is followed. The commutation properties are accounted for by appropriately defining matrix elements for infinitesimal time evolution as dictated by the Hamiltonian. We now show that the correlation function of the angular momentum operators is proportional to the same classical field correlations.

To do so we note that the angular momentum in its eigenbasis $\left| m\right>$ is given by

\begin{equation}\label{Lzbasis}
{\bf L}_{z} = \sum_{m} m \left| m\right>\left<m\right|
\end{equation}

\noindent We can now look at the operation on a state $\left|\theta\right>$ at site $i$ and time $\tau$,

\begin{eqnarray}\label{Lzbasis1}
{\bf L}_{z i \tau}\left|\theta_{i\tau}\right> &=& \sum_{m_{i\tau}} m_{i\tau} \left| m_{i\tau}\right>\left< m\right|\left. \theta_{i\tau}\right> \nonumber\\ 
&=& \sum_{m_{i\tau}} m_{i\tau} \left| m_{i\tau}\right>e^{\imath\theta_{i\tau}}
\end{eqnarray}

\noindent The correlation of the angular momentum operator in the theta basis is

\begin{eqnarray}
& &\left<\theta_{i\tau}\right| {\bf L}_{z i \tau }^{\dag}{\bf L}_{z j \tau '}\left|\theta_{j\tau '}\right> \\ \nonumber  &=& \sum_{m_{i\tau},m_{j\tau '}} m_{i\tau} m_{j\tau '} e^{-\imath m_{i\tau}\theta_{i\tau}+\imath m_{j\tau '}\theta_{j\tau '}}\left< m_{i\tau} \right|\left. m_{j\tau '}\right>\nonumber 
\end{eqnarray}

\begin{figure}[ht!]
  \centerline{
  \includegraphics[width=0.9\columnwidth]{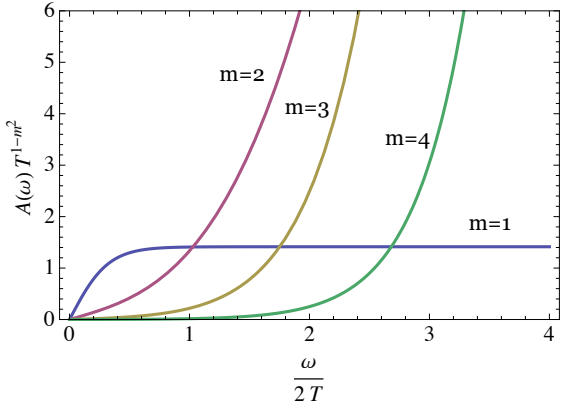}}
  \caption{The spectral function of the autocrrelation as a function of , shown. For m's larger than 1 the spectral weight shifts to higher frequency leading to short range correlation in time. For m=1 the spectrum leads to a power law decay.}
   \label{higherm}
\end{figure}

\noindent  Since the $\left<\exp(\imath\theta({\bf r},t))\exp(\imath\theta({\bf r}',t'))\right>$ correlations are local the same is assumed to be true for the angular momentum. This is justified because in the critical regime the spin waves are not the critical modes and the overlap of the angular momentum eigenstates $\left< m_{i\tau} \right|\left. m_{j\tau '}\right>$ which are spatially separated falls off exponentially. At the same site $\left< m_{i\tau} \right|\left. m_{i\tau '}\right> = \delta_{m_{i\tau}m_{i\tau '}}$. Having obtained the representation of the angular momentum correlation in the $\theta$ basis, we can compute the correlation function.

\begin{eqnarray}\label{corrsum}
C^{L_{z}}_{i,i,\tau,\tau '} &=& \left< {\bf L}_{z i \tau}^{\dag}{\bf L}_{z i \tau'}\right>  \\ \nonumber &=&  \left< \sum_{m_{i\tau},m_{i\tau '}} \delta_{m_{i\tau}m_{i\tau '}}m_{i\tau}m_{j\tau '}e^{-\imath m_{i\tau}\theta_{i\tau}+\imath m_{j\tau '}\theta_{j\tau '}}
\right>
\end{eqnarray}

\noindent  The $m_{i\tau}=1$ and $m_{j\tau '}=1$ contribution to the angular momentum correlations in the sum in eqn.\ref{corrsum} is equal to $\left<\exp(\imath\theta({\bf r},t))\exp(\imath\theta({\bf r}',t'))\right>$. Thus the leading term is local in space and power law in time. Higher order in $m$ correlations are also local in space but decay much faster in time so that the dominant contribution to ${\bf L}_{z}^{\dag}{\bf L}_{z}$ correlations is precisely of the form in eqn.\ref{eq:flucspec}. We refer to the derivation in ref.\cite{aji-cmv/warps} for further details but quote the main result. The correlations are given by 

\begin{equation}
\left<\exp(\imath m_{\textbf{r},\tau}\theta({\bf r},\tau))\exp(\imath m_{\textbf{r}',\tau '}\theta({\bf r}',\tau '))\right>  \propto  \delta_{\textbf{r},\textbf{r}'} \exp^{F\left(\tau-\tau'\right)} 
\end{equation}

\noindent and 

\begin{equation}
F\left(\tau-\tau'\right) =  - 2\pi T m^{2}{{1-\cos\left(\omega_{n}\left(\tau-\tau'\right)\right)}\over {\left|\omega_{n}\right|}}\log\left(\omega_{n}\tau_{c}\right)
\end{equation}

\noindent where $\omega_{n}$ is the matsubara frequency and $\tau_{c}$ is the short time cutoff in the theory. For $m=1$, the long time correlations decay as $\left|\tau-\tau '\right|^{-1}$. For $m > 1$, the correlations decay faster as the weight is shifted to higher and higher frequencies. One can infer this from the spectral function associated with this autocorrelator which is given by \cite{pouyan}

\begin{equation}
A_{\omega} = c_{0}T^{m^{2}-1}{{\sinh\left({\omega\over {2 T}}\right)\left|\Gamma\left({m^{2}\over 2} - {\imath\omega\over{2\pi T}}\right)\right|^{2}}\over{2^{\left(m^{2}-2\right)/2}\sqrt{\pi}\Gamma\left({m^{2}\over 2}\right)\Gamma\left({{m^{2}+1}\over 2}\right)}}
\end{equation}
\noindent where $c_{0}$ is a constant. To show the shift in weight we plot the spectral function for $m=1,2,3$ and $4$ in fig.\ref{higherm}. As $m$ becomes larger the frequency dependence at low frequencies is superlinear. The increasing weight at larger frequencies implies that the these correlations will decay faster that the $m=1$ contribution.

\end{document}